\documentclass[a4paper,aps,preprintnumbers,twocolumn,amsmath,amssymb,prl]{revtex4}
\usepackage{mathrsfs}
\usepackage{amsmath}
\usepackage[dvips]{graphicx}
\usepackage{epsfig}
\usepackage[dvips]{graphics,graphicx}

\begin{document}

\title{Probing superfluidity of periodically trapped ultracold atoms in a cavity by transmission spectroscopy }

\author{Aranya B.\ Bhattacherjee$^{1,2}$, Tarun Kumar$^{3}$ and ManMohan$^{3}$}

\address{$^{1}$Department of Physics, ARSD College, University of Delhi (South Campus), New Delhi-110021, India}\address{$^{2}$ Max Planck-Institute f\"ur Physik komplexer Systeme, N\"othnitzer Str.38, 01187 Dresden, Germany } \address{$^{3}$Department of Physics and Astrophysics, University of Delhi, Delhi-110007, India}

\begin{abstract}
We study a system of periodic Bose condensed atoms coupled to cavity photons using the input-output formalism of \cite{Collett}. We show that the cavity will either act as a through pass Lorentzian filter when the superfluid fraction of the condensate is minimum or completely reflect the input field when the superfluid fraction is maximum. We show that by monitoring the ratio between the transmitted field and the reflected field, one can estimate the superfluid fraction.
\end{abstract}


\maketitle

Experimental implementation of a combination of cold atoms and cavity QED (quantum electrodynamics) has made significant progress \cite{Nagorny03,Sauer04,Anton05}. This is a conceptually new regime of cavity QED, in which all atoms occupy a single mode of a matter-wave field and couple identically to the cavity light field, sharing a single excitation. It has been shown theoretically that the strong coupling of the condensed atoms to the cavity mode changes the resonance frequency of the cavity \cite{Horak00} and finite cavity response times lead to damping of the coupled atom-field excitations \cite{Horak01}. The driving field in the cavity can significantly enhance the localization and the cooling properties of the system\cite{Griessner04,Maschler04}. It has been demonstrated that in a cavity the atomic back action on the field introduces atom-field entanglement which modifies the associated quantum phase transition \cite{Maschler05}. The light field and the atoms become strongly entangled if the latter are in a superfluid state, in which case the photon statistics typically exhibits complicated multimodal structures \cite{Chen07}.  Because of the strong coupling of the condensate wave function to the cavity modes, a band structure of the condensate also leads to a band structure of the intracavity light fields. This in turn influences the Bloch energies, effective mass, Bogoliubov excitations and the superfluid fraction of the BEC \cite{Bhattacherjee}. 

Scattering of light from different atomic quantum states creates different quantum states of the scattered light, which can be distinguished by measuring the photon statistics of the transmitted light. For the Mott insulator, the number of photons scattered is zero while for the superfluid state it is nonzero and proportional to the number of atoms \cite{Mekhov07, Mekprl07,Mekpra07}. However one would also be interested in measuring the superfluid fraction of the Bose-Einstein condensate (BEC) which forms the motivation for the present work. Following the work of Collett and Gardiner \cite{Collett}, we link the internal scattered light with the external input and the output field using the boundary conditions at the cavity mirrors and show that the output field can be manipulated coherently by the state of the BEC and the external pump.

We consider an elongated cigar shaped Bose-Einstein condensate of $N$ two-level $^{87} Rb$ atoms in the $|F=1>$ state with mass $m$ and frequency $\omega_{a}$ of the $|F=1>\rightarrow |F'=2>$ transition of the $D_{2}$ line of $^{87} Rb$, strongly interacting with a quantized single standing wave cavity mode of frequency $\omega_{c}$. The cavity field is also coupled to external fields incident form the two side mirrors. The internal field is linked with the input by identification of the noise with the incoming field and the output can then be calculated using the boundary condition at the cavity mirror. In order to create an elongated BEC, the frequency of the harmonic trap along the transverse direction should be much larger than one in the axial (along the direction of the optical lattice) direction. The system is also coherently driven by a laser field with frequency $\omega_{p}$ through the cavity mirror with amplitude $\eta$. The two sided cavity has two partially transparent mirrors with associated loss coefficient $\gamma_{1}$ and $\gamma_{2}$. It is well known that high-Q optical cavities can significantly isolate the system from its environment, thus strongly reducing decoherence and ensuring that the light field remains quantum-mechanical for the duration of the experiment. The harmonic confinement along the directions perpendicular to the optical lattice is taken to be large so that the system effectively reduces to one-dimension. This system is modeled by a Jaynes-Cummings type of Hamiltonian $(H_{JC})$ in a rotating wave and dipole approximation \cite{Maschler05}

\begin{eqnarray}
H_{JC}&=&\dfrac{p^2}{2m}-\hbar \Delta_{a} \sigma^{+} \sigma^{-} -\hbar \Delta_{c}\hat{a}^{\dagger}\hat{a}\nonumber \\&-&i\hbar g(x)\left[ \sigma^{+}\hat{a}-\sigma^{-}\hat{a}^{\dagger}\right]-i\eta(\hat{a}-\hat{a}^{\dagger})\;
\end{eqnarray}

where $\Delta_{a}=\omega_{p}-\omega_{a}$ and $\Delta_{c}=\omega_{p}-\omega_{c}$ are the large atom-pump and cavity-pump detuning, respectively and  $\Delta_{c}>\Delta_{a}$. In this work we will consider only the case  $\Delta_{a}>0$. Here $\sigma^{+} , \sigma^{-}$ are the Pauli matrices. The atom-field coupling is written as $g(x)=g_{0} \cos(kx)$. Here $\hat{a}$ is the annihilation operator for a cavity photon. Since the detuning $\Delta_{a}$ is large, spontaneous emission is negligible and we can adiabatically eliminate the excited state using the Heisenberg equation of motion $\dot{\sigma^{-}}=\dfrac{i}{\hbar}\left[ H_{JC},\sigma^{-}\right] $. This yields the single particle Hamiltonian

\begin{equation}
H_{0}=\dfrac{p^2}{2m}-\hbar \Delta_{c}\hat{a}^{\dagger}\hat{a}+\hbar U_{0} \cos^2(kx)\left[ 1+\hat{a}^{\dagger} \hat{a}\right]-i\eta(\hat{a}-\hat{a}^{\dagger}).
\end{equation}

The parameter $U_{0}=\dfrac{g_{0}^{2}}{\Delta_{a}}$ is the optical lattice barrier height per photon and represents the atomic backaction on the field \cite{Maschler05}. Here we will always take $U_{0}>0$. In this case the condensate is attracted to the nodes of the light field and hence the lowest bound state is localized at these positions which leads to a reduced coupling of the condensate to the cavity compared to that for $U_{0}<0$.  Along $x$, the cavity field forms an optical lattice potential of period $\lambda/2$ and depth $\hbar U_{0}(\hat{a}^{\dagger}\hat{a}+1)$. We now write the Hamiltonian in a second quantized form including the two body interaction term.

\begin{eqnarray}
H&=&\int d^3 x \Psi^{\dagger}(\vec{r})H_{0}\Psi(\vec{r})\nonumber \\&+&\dfrac{1}{2}\dfrac{4\pi a_{s}\hbar^{2}}{m}\int d^3 x \Psi^{\dagger}(\vec{r})\Psi^{\dagger}(\vec{r})\Psi(\vec{r})\Psi(\vec{r})\;
\end{eqnarray}

where $\Psi(\vec{r})$ is the field operator for the atoms. Here $a_{s}$ is the two body $s$-wave scattering length. The corresponding Bose-Hubbard Hamiltonian can be derived by writing $\Psi(\vec{r})=\sum_{i} \hat{b}_{i} w(\vec{r}-\vec{r}_{i})$, where $w(\vec{r}-\vec{r}_{i})$ is the Wannier function and $\hat{b}_{i}$ is the corresponding annihilation operator for the bosonic atom. Retaining only the lowest band with nearest neighbor interaction, we have

\begin{eqnarray}
H &=& E_{0}\sum_{j}\hat{b}_{j}^{\dagger}\hat{b}_{j}+E\sum_{j}\left(\hat{b}_{j+1}^{\dagger}\hat{b}_{j}+\hat{b}_{j+1}\hat{b}_{j}^{\dagger} \right)\nonumber \\&+&\hbar U_{0}(\hat{a}^{\dagger}\hat{a}+1)\left\lbrace J_{0}\sum_{j}\hat{b}_{j}^{\dagger}\hat{b}_{j}+J \sum_{j}\left(\hat{b}_{j+1}^{\dagger}\hat{b}_{j}+\hat{b}_{j+1}\hat{b}_{j}^{\dagger} \right)\right\rbrace\nonumber \\&-&\hbar \Delta_{c} \hat{a}^{\dagger}\hat{a}-i\hbar \eta (\hat{a}-\hat{a}^{\dagger})+\dfrac{U}{2}\sum_{j}\hat{b}_{j}^{\dagger}\hat{b}_{j}^{\dagger}\hat{b}_{j}\hat{b}_{j}\;
\end{eqnarray}

where

\begin{eqnarray}
U&=&\dfrac{4\pi a_{s}\hbar^{2}}{m}\int d^3 x|w(\vec{r})|^{4}\nonumber \\
E_{0}&=&\int d^3 x w(\vec{r}-\vec{r}_{j})\left( -\dfrac{\hbar^2 \nabla^{2}}{2m}\right)w(\vec{r}-\vec{r}_{j})\nonumber \\
E &=&\int d^3 x w(\vec{r}-\vec{r}_{j})\left( -\dfrac{\hbar^2 \nabla^{2}}{2m}\right)w(\vec{r}-\vec{r}_{j \pm 1})\nonumber \\
J_{0}&=&\int d^3 x w(\vec{r}-\vec{r}_{j}) \cos^2(kx)w(\vec{r}-\vec{r}_{j})\nonumber \\
J &=&\int d^3 x w(\vec{r}-\vec{r}_{j}) \cos^2(kx)w(\vec{r}-\vec{r}_{j \pm 1}).
\end{eqnarray}

The nearest neighbor nonlinear interaction terms are usually very small compared to the onsite interaction and are neglected as usual. The onsite energies $J_{0}$ and $E_{0}$ are set to zero. We now write down the Heisenberg equation of motion for the bosonic field operator $\hat{b}$ as

\begin{eqnarray}
\dot{\hat{b}}_{j}&=&-iU_{0}\left( 1+\hat{a}^{\dagger} \hat{a} \right)J\left\lbrace \hat{b}_{j+1}+\hat{b}_{j-1} \right\rbrace-\dfrac{iE}{\hbar}\left\lbrace\hat{b}_{j+1}+\hat{b}_{j-1}  \right\rbrace\nonumber \\&-& \dfrac{iUn_{0}}{\hbar}\hat{b}_{j}\;
\end{eqnarray}

The behaviour of the internal cavity mode is obtained from the quantum-Langevin equation which for a single-mode cavity becomes

\begin{eqnarray}
\dot{\hat{a}}&=&-iU_{0}\left\lbrace J_{0}\sum_{j}\hat{b}_{j}^{\dagger}\hat{b}_{j}+J \sum_{j} \left(\hat{b}_{j+1}^{\dagger}\hat{b}_{j}+\hat{b}_{j+1}\hat{b}_{j}^{\dagger} \right)\right\rbrace \hat{a}+\eta\nonumber \\&+&i\Delta_{c} \hat{a}-\dfrac{\gamma_{1}}{2} \hat{a}-\dfrac{\gamma_{2}}{2} \hat{a}+\sqrt{\gamma_{1}}\hat{a}_{in}+\sqrt{\gamma_{2}}\hat{b}_{in}\;
\end{eqnarray}

Here $\hat{a}_{in}$ and $\hat{b}_{in}$ are the external input fields incident from the two mirrors.. Equation (6) and (7) represents a set of coupled equations describing the dynamics of the compound system formed by the condensate and the optical cavity. We will work in the bad cavity limit, where typically, $\gamma_{1}$ and $\gamma_{2}$ are the fastest time scale ( this means that the cavity decay rates are much larger than the oscillation frequency of bound atoms in the optical lattice of the cavity ). In this limit the intracavity field adiabatically follows the condensate wavefunction, and hence we can put $\dot{\hat{a}}=0$. We treat the BEC within the mean field framework (large atom numbers) and assume the tight binding approximation where we replace $\hat{b}_{j}$ by $\phi_{j}$ and look for solutions in the form of Bloch waves

\begin{equation}
\phi_{j}=u_{k}exp(ikjd)exp(-i\mu t/ \hbar).
\end{equation}

Here $\mu$ is the chemical potential, $d$ is the periodicity of the lattice and $\dfrac{1}{I}\sum_{j}\hat{b}_{j}^{\dagger} \hat{b}_{j}=|u_{k}|^{2}=n_{0}$ (atomic number density)and $I$ is the total number of lattice sites. Also $\sum_{j} n_{0}=N$ (total number of atoms) In frequency space we obtain from equations (6) and (7)

\begin{equation}
\tilde{a}(\omega)=\dfrac{\eta+ \sqrt{\gamma_{1}}\tilde{a}_{in}(\omega)+ \sqrt{\gamma_{2}}\tilde{b}_{in}(\omega)}{\left\lbrace \gamma_{1}/2+\gamma_{2}/2-i\left( \Delta_{c}+\omega-2JNU_{0}\cos(kd)\right) \right\rbrace }.
\end{equation}

Where,

\begin{equation}
\tilde{a}(\omega)=\dfrac{1}{\sqrt{2 \pi}} \int_{-\infty}^{\infty} e^{i \omega t} \hat{a}(t) dt.
\end{equation}

The fields $\hat{a}_{in}$ and $\hat{b}_{in}$ are also related to their frequency components in a similar manner. Interestingly we find that due to the atomic backaction, the quantum state of the cavity field varies along the Brillioun zone. The cavity photons develops a band structure due to the strong coupling with the condensate, analogous to photonic band gap materials. The concept of photonic band gaps in optical lattices has been known for quite some time \cite{Deutsch94}. The cavity photons are created by scattering through the atoms which are coherently driven by the pump. Interestingly, the average photon number $<\hat a^{\dagger} \hat a>$ measures the light transmission spectra and is different for the Mott insulator (MI) and the superfluid phase (SF) \cite{Mekhov07}.

The relationship between the input and output modes may be found from using the boundary conditions at each mirror,

\begin{equation}
\tilde{a}_{out}(\omega)+\tilde{a}_{in}(\omega)=\sqrt{\gamma_{1}} \tilde{a}(\omega)
\end{equation}

\begin{equation}
\tilde{b}_{out}(\omega)+\tilde{b}_{in}(\omega)=\sqrt{\gamma_{2}} \tilde{a}(\omega)
\end{equation}

We find,

\begin{equation}
\tilde{a}_{out}(\omega)=\dfrac{\eta \sqrt{\gamma_{1}}+(\gamma_{1}/2-\gamma_{2}/2+i \Delta)\tilde{a}_{in}(\omega)+ \sqrt{\gamma_{1} \gamma_{2}}\tilde{b}_{in}(\omega)}{\left\lbrace \gamma_{1}/2+\gamma_{2}/2-i \Delta^{'} \right\rbrace }.
\end{equation}

Where $\Delta^{'}= \Delta_{c}+\omega-2JNU_{0}\cos(kd)$. If the two mirrors are the same, $\gamma_{1}=\gamma_{2}=\gamma$ and near resonance $\Delta^{'} \approx 0$. The resonance point is one where the superfluid fraction is minimum \cite{Bhattacherjee}.

\begin{equation}
\tilde{a}_{out}(\omega) \approx \dfrac{\sqrt{\gamma} \eta +\gamma \tilde{b}_{in}(\omega)}{\left\lbrace \gamma-i \Delta^{'} \right\rbrace }.
\end{equation}

This shows that the cavity now behaves like a shifted through-pass Lorentzian filter. The input field will be completely reflected if $\Delta^{'}>> \gamma$ (when the atoms are in the deep superfluid regime), $\tilde{a}_{out}(\omega)\approx - \tilde{a}_{in}(\omega)$. The state of the BEC together with the cavity parameters control the output field of the cavity. The parameter $\Delta_{c}-2JNU_{0}\cos(kd)$ controls the superfluid fraction \cite{Bhattacherjee}. By monitoring the ratio between the transmitted field and the reflected field, one can estimate the superfluid fraction. This is the main result of this paper.

In order to monitor the superfluidity, one can perform a transmission spectroscopy with the scattered light by direct read out of the number of photons coming out of the cavity. The transmission of the scattered light is monitored as a function of $\Delta$.  Photon loss can be minimized by using high-Q cavities and thus ensuring that the light field remains quantum-mechanical for the duration of the experiment.  Recent experiments \cite{Klinner06} showed that in a ring cavity, even at large detunings from the atomic resonance a strong coupling between the atoms and the cavity field can be achieved.
In conclusion, we have studied a system of periodic Bose condensed atoms coupled to cavity photons using the input-output formalism of \cite{Collett}. We have shown that the cavity will either act as a through pass Lorentzian filter when the superfluid fraction is minimum or completely reflect the input field when the superfluid fraction is maximum. 

\section{Acknowledgements}

One of the authors Tarun Kumar acknowledges the Council of University Grants Commission, New Delhi for the financial support under the Junior Research Fellowship scheme Sch/JRF/AA/30/2008-2009.

\end{document}